\documentclass[11pt,letterpaper]{article}
\usepackage{mystyle,epsfig}
\usepackage[T1]{fontenc}
\def\be{\begin{equation}}
\def\ee{\end{equation}}
\def\bea{\begin{eqnarray}}
\def\eea{\end{eqnarray}}

\begin{document}
\vspace*{4cm}
\title{SEARCH FOR THE ASSOCIATED PRODUCTION OF CHARGINO AND NEUTRALINO IN THE FINAL STATE WITH THREE LEPTONS}
\author{A.~Canepa for the CDF Collaboration}

\address{Department of Physics, Purdue University, 525 Northwestern Avenue,\\
West Lafayette, IN 47906, USA}


\maketitle
\abstracts{
Supersymmetry, if realized in Nature, predicts the existence of new particles, as chargino  
and neutralino, which might manifest themselves
with peculiar signatures. Three leptons and large missing transverse energy in the event could
signal their associated production. 
We report the latest results
of the search performed by the CDF Collaboration in $\sqrt{s}=1.96$ TeV proton-antiproton collisions at Tevatron Run II.
}


\section{Introduction}
The Standard Model (SM) of particle physics emerged in the 1970's and high-precision experiments have repeatedly verified
its predictions. Nevertheless the SM is regarded as a low energy effective theory because of 
experimental and theoretical unresolved issues. It explains the origin of mass by means of the Higgs mechanism 
but the Higgs boson has not been observed yet. It does not include gravity nor it provides gauge unification. It 
does not account for the Dark Matter of the Universe.
These open questions stimulate a rich search for physics beyond the SM. On theoretical grounds, the most promising 
extension 
of the SM is the Supersymmetric SM (SSM). Supersymmetry~\cite{martin} (SUSY) is a proposed symmetry of Nature
which introduces a fermion (boson) for each SM boson (fermion) with the same quantum numbers but the spin.  
A discrete multiplicative symmetry, called R-parity, is defined as $R_P=(-1)^{2S+3B+L}$ such that a SM particle 
carries $R_P=+1$ and a SUSY particle $R_P=-1$. Supersymmetric particles (sparticles) have not been observed yet implying that SUSY is a broken symmetry. 
The minimal supergravity model (mSUGRA) with R-parity conservation, is a favored   
breaking model for SUSY. In the mSUGRA scenario the sparticles are produced
in pairs and the lighter charginos and neutralinos (mixed state of electroweak gauginos and higgsinos)
and the sleptons,  are less massive than gluinos and squarks.
If SUSY is a broken symmetry, it predicts a low mass Higgs boson in accord
with the electroweak fits, accomodates gravity, unifies the gauge interactions and provides an excellent candidate 
for Dark Matter. In particular for the mSUGRA model, the lightest neutralino ${\tilde\chi}^{0}_{1}$ is identified as the candidate for Dark Matter, being neutral and the lightest stable sparticle (LSP).
In this paper we report on the search for the associated production of chargino and neutralino when these particles
decay leptonically into three charged leptons and two neutralinos ${\tilde\chi}^{0}_{1}$ which escape the
 detection causing a significant missing transverse energy in the event. This channel is reckoned as the Golden Mode 
for SUSY at a 
hadron collider. The analysis proceeds as a counting experiment by comparing the SM prediction to the 
observed data in kinematic regions where the SUSY signal is expected to be negligible (``control regions''). It 
is performed as a ``blind'' analysis. ``Blind'' analysis means that the region of the data where the SUSY signal is enhanced with respect to the SM 
background (``signal region'') is investigated only if the agreement between the expectation and the observation
is yielded in the control regions.

\section{Production and decay at the Tevatron}
The ${\tilde\chi}^{\pm}_{1}{\tilde\chi}^{0}_{2}$ associated production is expected to occur via two modes which
interfere destructively: a dominant s-mode, through virtual W exchange and a suppressed t-mode, through virtual 
squark exchange. The charginos and neutralinos can decay into charged leptons via virtual sleptons or virtual W/Z, as for
instance ${\tilde\chi}^{\pm}_{1}\rightarrow W^{*}{\tilde\chi}^{0}_{1} \rightarrow \ell\nu{\tilde\chi}^{0}_{1}$.  
The mSUGRA benchmark point selected for performing the analysis  
corresponds to a chargino mass at the boundary of the LEPII exclusion limit~\footnote{The mass of the chargino is 
$m_{\tilde{\chi}^{\pm}_{1}} = 113$ GeV/c$^{2}$. The corresponding mSUGRA parameters are the following:
$m_{\frac{1}{2}} = 180$ GeV/c$^{2} ;$ $m_{0} = 100 $ GeV/c$^{2}$; tan$\beta = 5$; $\mu > 0$; $A_{0}= 0 $.}.
The next-to-leading production cross section calculated with the algorithm Prospino~\cite{prospino} is 
$\sigma_{{\tilde\chi}^{\pm}_{1}{\tilde\chi}^{0}_{2}}$ = 0.642 pb. The fully leptonic branching ratio obtained with 
the Monte Carlo Event Generator PYTHIA~\cite{pythia} is 0.25. 

\section{Search at CDF}
\begin{figure}[ht]
\begin{minipage}[h]{8cm}
\begin{center}
\includegraphics[width=0.25\textheight]{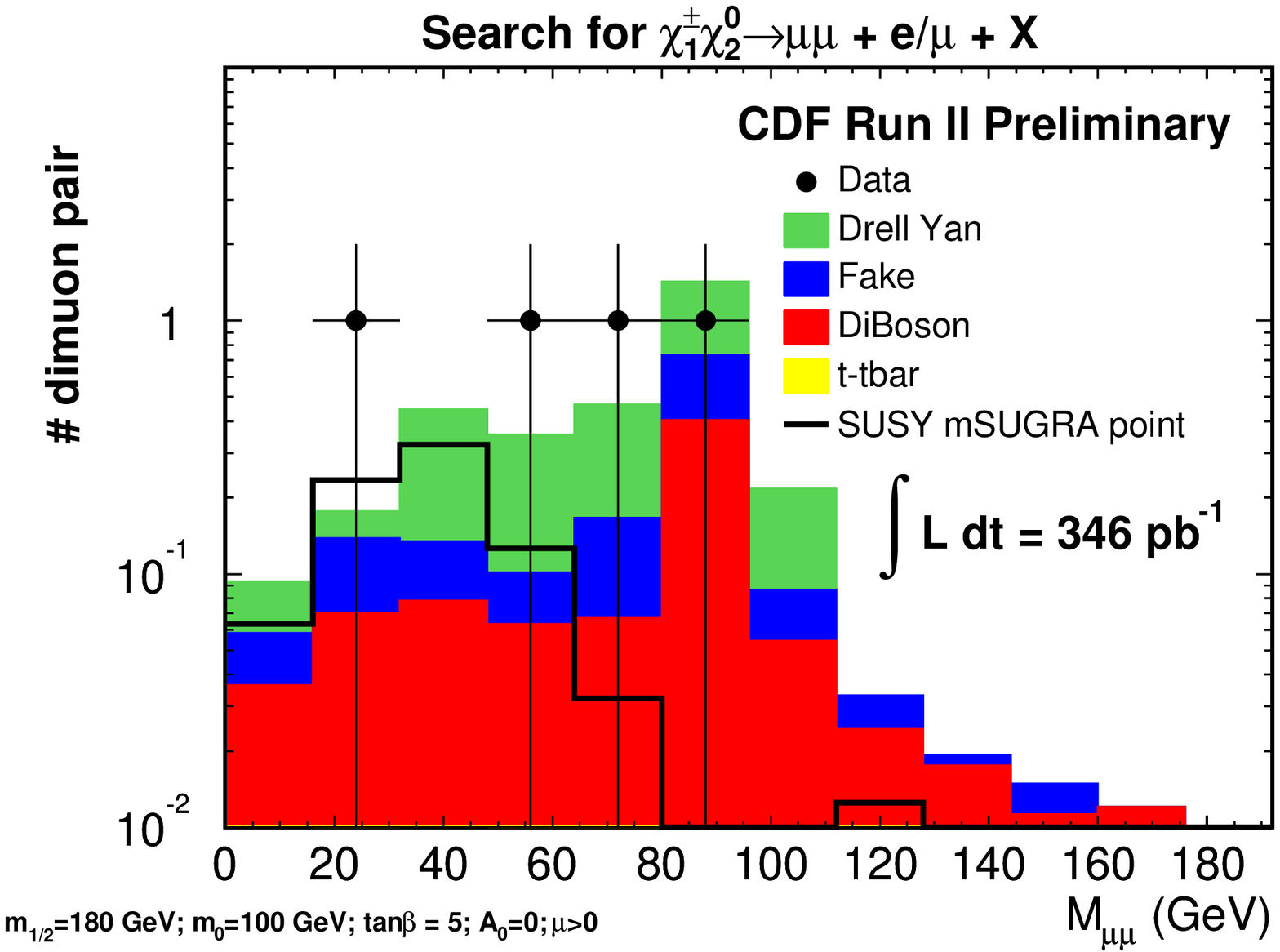} \caption{Invariant mass distribution of opposite sign muons in trilepton events ($\mu\mu+e/\mu$) \label{massAnadi}}
\end{center}
\end{minipage}
\hfill
\begin{minipage}[h]{7.5cm}
\includegraphics[width=0.75\textwidth]{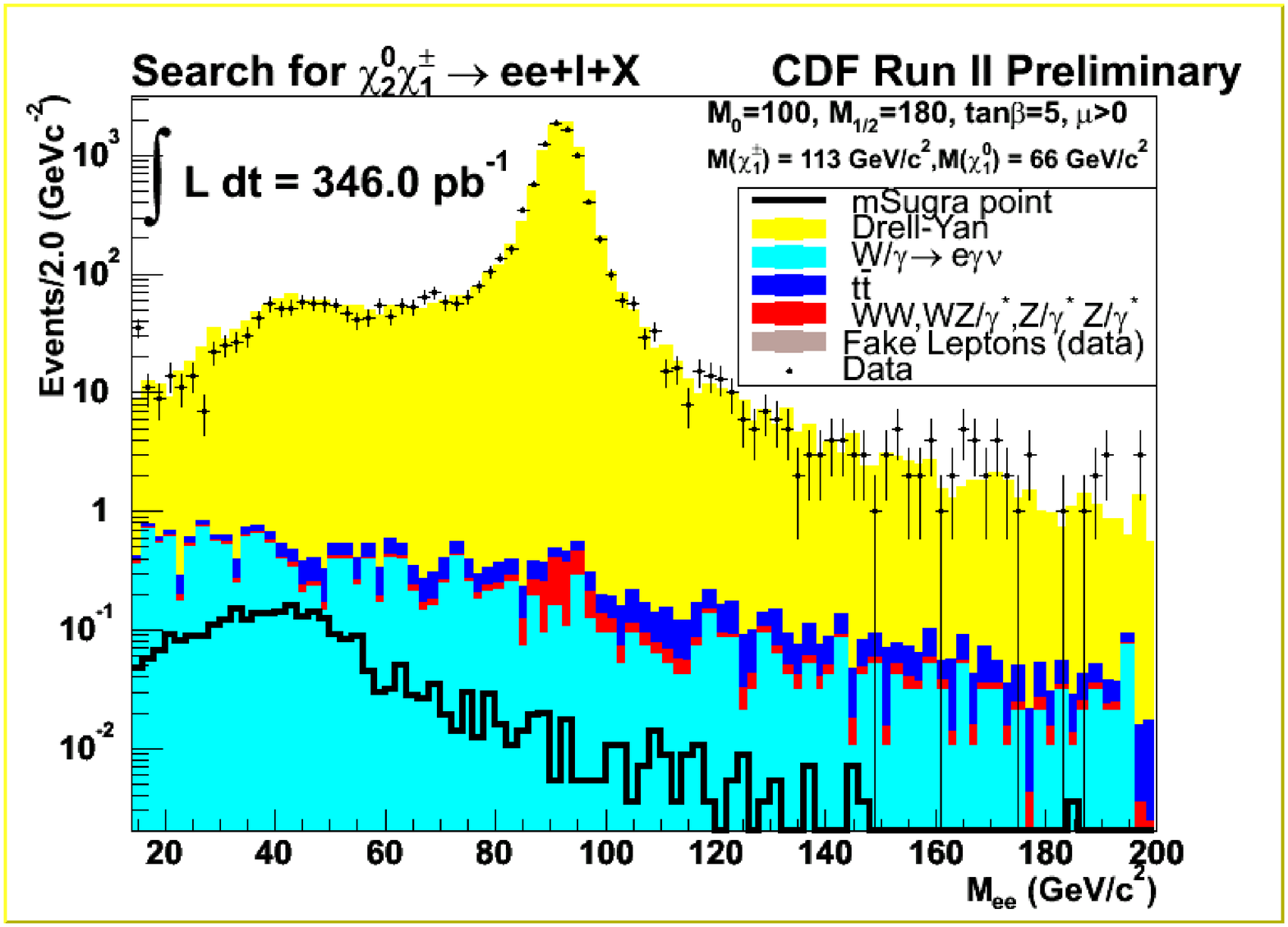} \caption{Invariant mass distribution of opposite sign electrons in dielectron events ($ee+\mu/e$) \label{massGiulia}}
\end{minipage}
\hfill
\end{figure}

\begin{figure}[ht]
\begin{minipage}[h]{8cm}
\begin{center}
\includegraphics[width=0.75\textwidth]{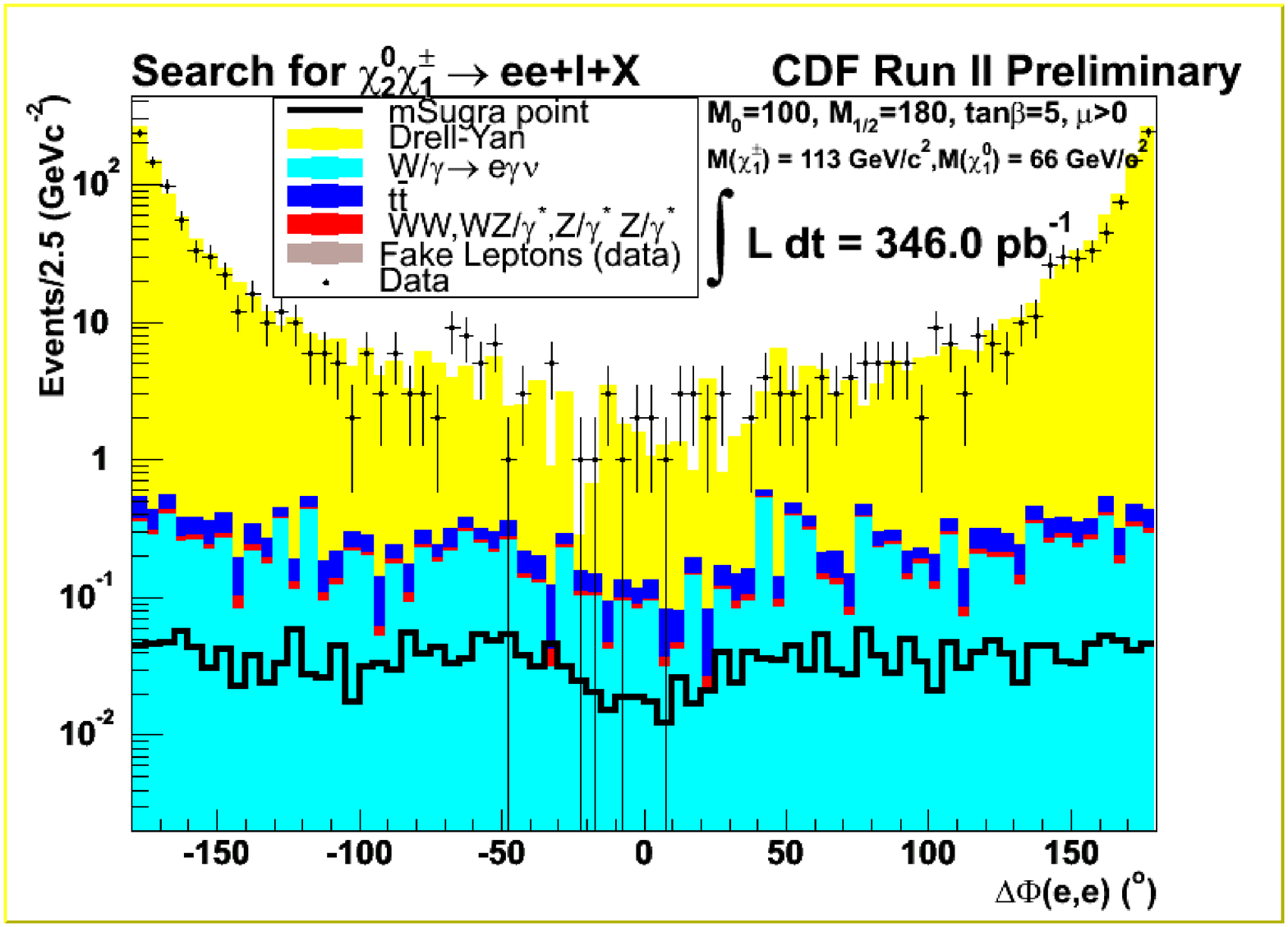} \caption{$\Delta\phi_{e^+e^-}$ in dielectron events with $M_{e^+e^-}$ in the allowed window ($ee+\mu/e$) \label{phiGiulia}}
\end{center}
\end{minipage}
\hfill
\begin{minipage}[h]{7.5cm}
\includegraphics[width=0.8\textwidth]{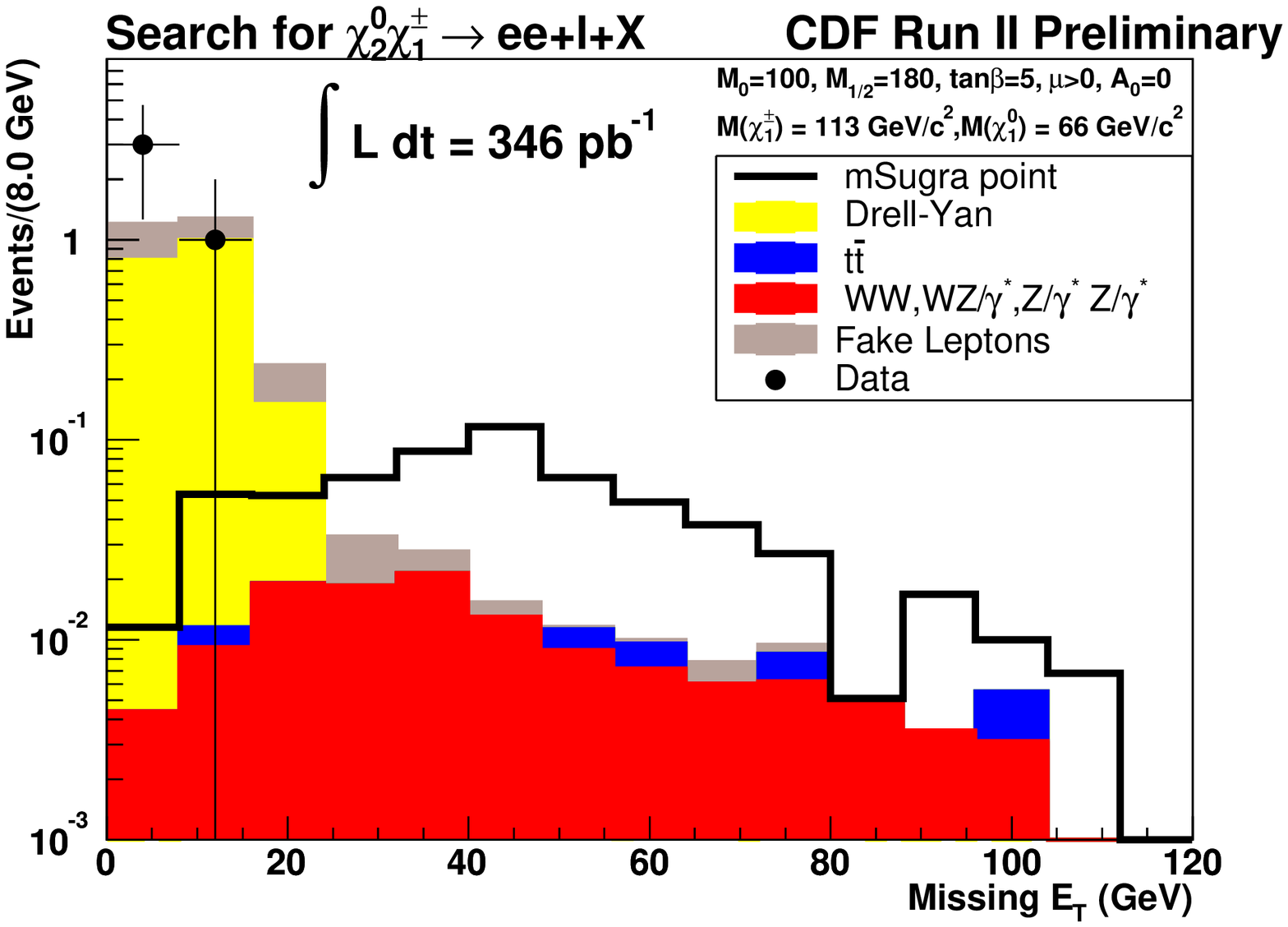} \caption{MET in the ``signal'' region ($ee+\mu/e$) \label{metGiulia3}}
\end{minipage}
\hfill
\end{figure}

\begin{figure}[ht]
\begin{minipage}[h]{8cm}
\begin{center}
\includegraphics[width=0.25\textheight]{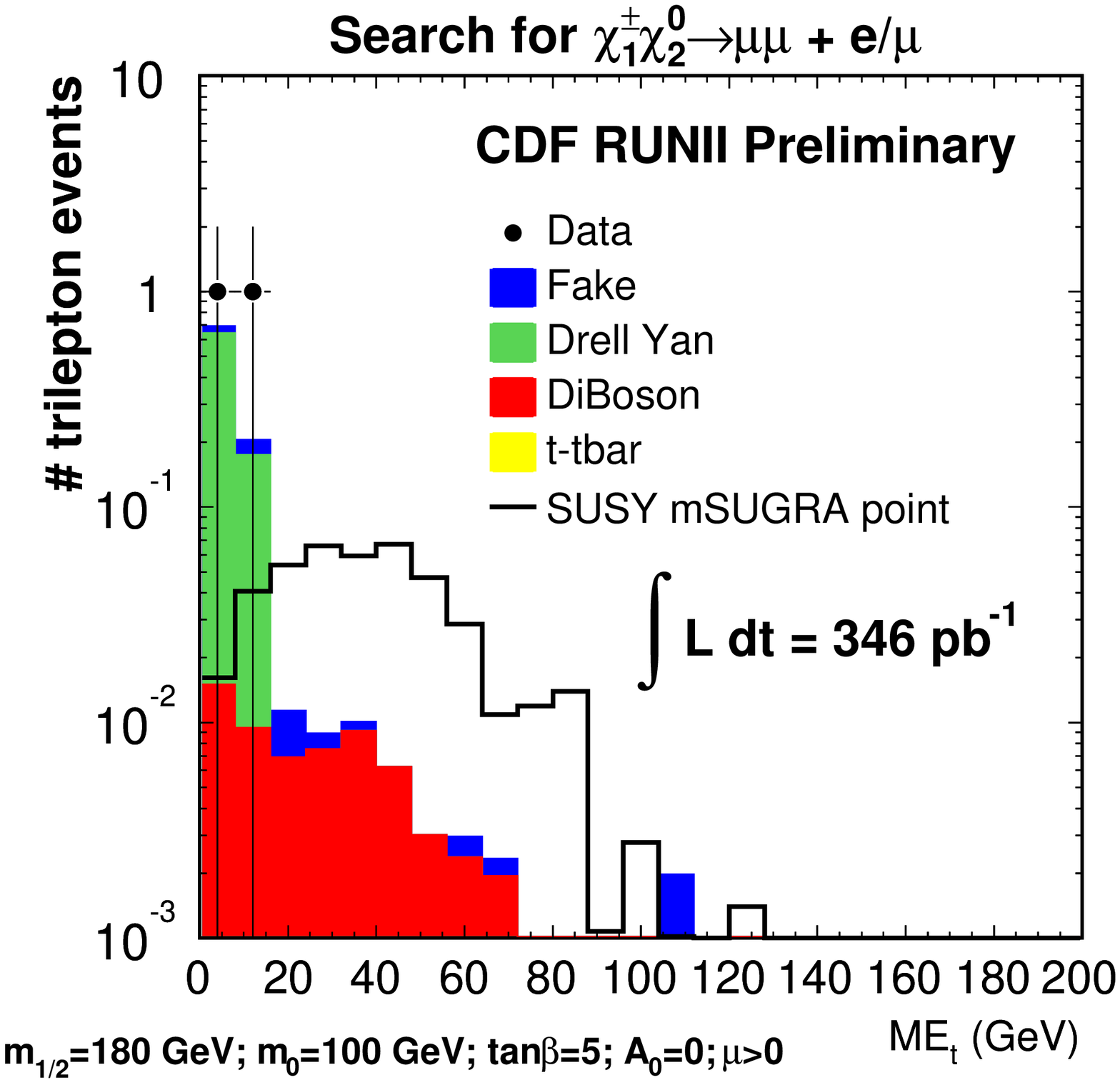} \caption{MET in the ``signal'' region ($\mu\mu+e/\mu$) \label{metAnadi}}
\end{center}
\end{minipage}
\hfill
\begin{minipage}[h]{7.5cm}
\begin{center}
\includegraphics[width=0.7\textwidth]{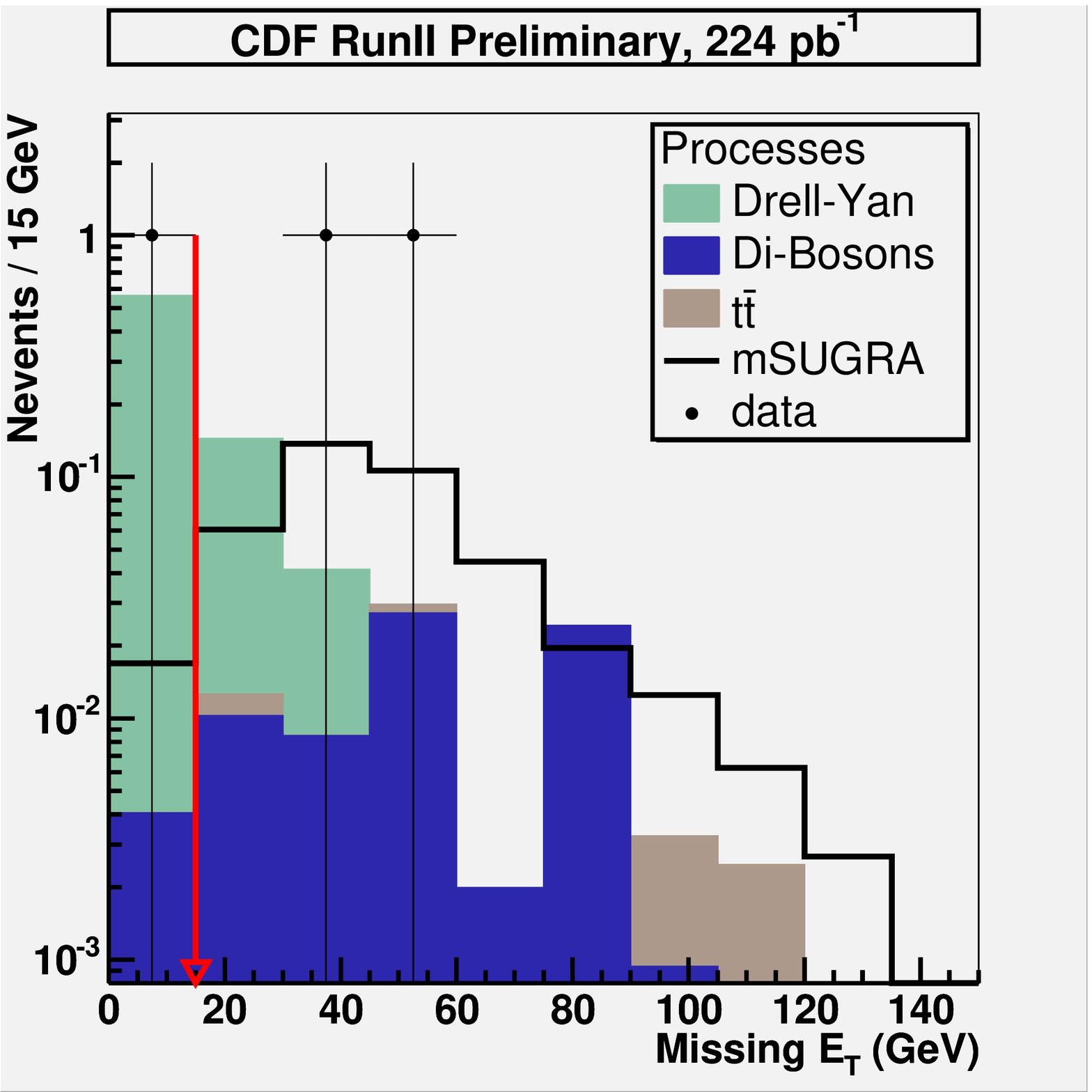} \caption{MET in the ``signal'' region ($ee+track$) \label{metJohn}}
\end{center}
\end{minipage}
\hfill
\end{figure}

The CDF~\cite{CDF} Collaboration is currently pursuing the search in three different channels for maximizing the 
acceptance: 
two analyses (referred as ``trilepton'' analyses) require three central 
leptons~\footnote{In the following we use lepton for electron or muon and MET for Missing Transverse 
Energy.}  ($|\eta|<1.0$) with transverse momentum $p_T >$ 20, 8 and 5 GeV/c; the highest $p_T$ leptons are 
constrained to be same flavor leptons. 
The third analysis (referred as `dielectron+track' analysis) is based on two central electrons accompanied by an  isolated track.
The trilepton analyses investigate the CDF dataset collected via the single lepton trigger (lepton $p_T >$ 18 GeV/c),
which grants high acceptance to decays of massive supersymmetric particles leading to high $p_T$ leptons.
The dielectron+track analysis is sensitive to leptons from $\tau$ decay which typically populate the low p$_{T}$ region;
therefore, the dataset collected via the dielectron trigger (electron $p_T >$ 4 GeV/c) is the most appropriate. 
Several SM processes yield the same signature as the SUSY signal, that is three spatially separated leptons
and significant MET. The SM backgrounds may depend on the channel under investigation 
($\mu\mu+e/\mu$, $ee+\mu/e$, $ee+track$) and specific selection criteria are applied accordingly.
The major background derives from Drell Yan production where the third lepton in the event originates from photon 
conversions or from hadrons misidentified as leptons; the leading leptons, distributed at high azimuthal angles, 
are expected to exhibit large p$_{T}$ and high invariant mass. 
The MET in the event may be significant and aligned to any of the leptons due to detector resolution effects.
Diboson events constitute a minor irreducible background whereas the smallest background is due to fully leptonic 
decay of heavy flavor. $t\bar{t}$ and $b\bar{b}$/$c\bar{c}$ events are 
discriminated from the SUSY signal for the large jet activity; in particular for $b\bar{b}$/$c\bar{c}$ 
production, the leptons carry low p$_{T}$ and appear non isolated in the calorimeter. 
The SM processes listed above (except for the Drell Yan production associated to a misidentified
lepton) are generated with PYTHIA and MADGRAPH~\cite{MAD} Monte Carlo event generators and processed through the full 
detector simulation to measure the acceptance. 
The estimation of the Drell Yan production associated to a misidentified lepton depends on the analysis. 
In case of the trilepton analyses, the misidentification probability is measured in a dataset depleted of
 real leptons and applied to the dataset of interest. 
In the dieletron+track analysis, the misidentification probability of the isolated track is extracted from Z events in data and 
used to weight the Drell Yan Monte Carlo to emulate the contribution on a wide mass spectrum. 
The selection criteria~\footnote{The selection criteria are optimized to gain the highest reach defined as $\frac{S}{\sqrt{B}}$.} 
applied to suppress the SM background are summarized in Table \ref{cuts}. 
\begin{table}
\begin{small}
\begin{center}
\begin{tabular}{||l|l|l||} \hline\hline
\ Selection Criteria                                               & SM background               & Analysis \\ \hline\hline
\ $15 < m_{\mu\mu} < 76$ GeV/c$^2$,  $m_{\mu\mu} > 106$ GeV/c$^2$  & Resonances (J$/\Psi$, $\Upsilon$, Z)    & $\mu\mu+e/\mu$, $ee+\mu/e$, $ee+track$ \\ 
\ $\Delta\phi_{ee} < 2.9$ rad                                      & Drell Yan                               & $ee+\mu/e$, $ee+track$  \\ 
\ MET $> 15$    GeV                                        & Drell Yan                               & $\mu\mu+e/\mu$, $ee+\mu/e$, $ee+track$ \\ 
\ $N_{jet} < 2$, jet $E_T>20$ GeV                          & $t\bar{t}$, $b\bar{b}$/$c\bar{c}$       & $\mu\mu+e/\mu$, $ee+\mu/e$  \\
\ $\sum_{jets}E_T <$ 80 GeV, jet $E_T>20$ GeV      & $t\bar{t}$, $b\bar{b}$/$c\bar{c}$       & $ee+track$ \\
\ $m_{T}^{e,MET} >$ 10 GeV                                 &Instrumental background                  & $ee+track$  \\ 
\hline\hline  
\end{tabular}
\caption{Event selection criteria (left column) applied to suppress the SM background (middle column) 
 for the different analyses (right column).\label{cuts}}
\end{center}
\end{small}
\end{table}
In terms on these, up to seventeen control regions are assigned and each control region
is mainly dominated by a single SM process. We observe good agreement between prediction and data in the control regions,
as shown in Figure \ref{massAnadi} to Figure \ref{metJohn}. The signal region, defined as the set of events
 satisfying the selection in Table \ref{cuts}, is therefore explored. The main systematic uncertainties can be divided
into uncertainties on the SUSY signal, uncertainties on the background and uncertainties common to both. As far
as the background is concerned, the largest systematic uncertainties in the electron and muon channels 
are respectively the uncertainty on the jet energy scale ($30\%$) and the uncertainty on the muon transverse
momentum (7$\%$). An additional 5$\%$ uncertainty in the trilepton analyses derives from the lepton misidentification probability. The largest systematic uncertainty on the SUSY signal prediction comes from the theoretical calculation
of the production cross section (7$\%$). The 6$\%$ uncertainty from the measurement of the integrated luminosity affects
both the SUSY signal prediction and the SM background prediction. The total uncertainty on the signal (background) acceptance 
adds up to 11$\%$ (12$\%$) and 8$\%$ (26$\%$) in the $\mu\mu+e/\mu$ and $ee+\mu/e$
 channel respectively. In the dielectron+track analysis, the total statistical and systematic uncertainty is 
14$\%$ (75$\%$) for the signal (background) prediction. Both electron channels suffer a large systematic uncertainty on the 
jet energy scale which is less relevant for the muon channel as events with jets are mainly rejected by a preselection cut
 ($20<\Delta\phi_{MET-jet}<160^{\circ} $). In the trilepton analyses we expect 0.09$\pm$0.03 \begin{tiny}STAT+SYS\end{tiny} SM 
events in the muon channel and 
0.17$\pm$0.05 \begin{tiny}STAT+SYS\end{tiny} SM events in the electron channel. In the dielectron+track analysis we expect 
0.36$\pm$0.27 \begin{tiny}STAT+SYS\end{tiny} SM events. The estimated number of events from the SUSY 
signal is 0.37$\pm$0.05 \begin{tiny}STAT+SYS\end{tiny} and 0.49$\pm$0.05 \begin{tiny}STAT+SYS\end{tiny} for the muon 
and electron trilepton analysis, and 0.36$\pm$0.27 \begin{tiny}STAT+SYS\end{tiny} for the dielectron+track analysis. 
These quotes refer to a luminosity of 346 pb$^{-1}$ for the trilepton analyses and 224  pb$^{-1}$ for the dielectron+track 
analysis, as shown in Tab. \ref{triResA} to Tab. \ref{triResJ}. In the trilepton analyses we observe 0 events in both the $\mu\mu+e/\mu$ and the $ee+\mu/e$ channel. 
In the dielectron+track analysis we observe 2 events.
\begin{table}[ht]
\begin{tiny}
\begin{center}
\begin{tabular}{||l|l||} \hline\hline
\  Process         & Number of events (346 pb$^{-1}$) \\ \hline\hline
\  Diboson         &$0.043\pm 0.003$  \\ 
\  Heavy Flavor    &$0.0 + 0.004$   \\ 
\  Fake Leptons    &$0.010\pm 0.005$   \\ 
\  Drell Yan       &$0.035\pm 0.035$   \\ 
\  TOTAL BACKGROUND& $0.09\pm 0.03$  \\ 
\  SUSY            &  $0.37\pm 0.03$   \\ 
\  DATA            &  $0$   \\ 
\hline\hline
\end{tabular}
\caption{Expected number of signal, background and observed events in 346 pb$^{-1}$ for the trilepton $\mu\mu+\mu/e$ analysis. The uncertainty includes statistical uncertainty.\label{triResA}}
\end{center}
\end{tiny}
\end{table}
\begin{table}[ht]
\begin{tiny}
\begin{center}
\begin{tabular}{||l|l||} \hline\hline
\  Process         & Number of events (346 pb$^{-1}$) \\ \hline\hline
\  Diboson         &$0.065\pm0.013$  \\ 
\  Heavy Flavor    &$0.007\pm0.004$   \\ 
\  Fake Leptons    &$0.031\pm0.015$   \\
\  Drell Yan       &$0.067\pm0.035$   \\ 
\  TOTAL BACKGROUND& $0.171\pm0.052$  \\
\  SUSY            &  $0.491\pm0.050$   \\ 
\  DATA            &  $0$   \\ 
\hline\hline
\end{tabular}
\caption{Expected number of signal, background and observed events in 346 pb$^{-1}$ for the trilepton $ee+\mu/e$ analysis. The total uncertainty is quoted.\label{triResG}}
\end{center}
\end{tiny}
\end{table}
\begin{table}[!tbp]
\begin{tiny}
\begin{center}
\begin{tabular}{||l|l||} \hline\hline
\  Process         & Number of events (224 pb$^{-1}$) \\ \hline\hline
\  WW + ZZ/Z$\gamma$         &$0.062\pm0.023$  \\
\  WZ/W$\gamma$      &$0.032\pm0.005$  \\ 
\  $t\bar{t}$       &$0.010\pm0.007$  \\
\  Heavy Flavor    &$0.0 + 0.21$   \\ 
\  Drell Yan       &$0.25\pm0.17$   \\ 
\  TOTAL BACKGROUND& $0.36\pm0.27$  \\ 
\  SUSY            &  $0.480\pm0.066$   \\ 
\  DATA            &  $2$   \\ 
\hline\hline
\end{tabular}
\caption{Expected number of signal, background and observed events in 224 pb$^{-1}$ for the dielectron analysis. The total uncertainty is quoted. \label{triResJ}}
\end{center}
\end{tiny}
\end{table}

\newpage
\section{Conclusions}
A search for the associated production of chargino and neutralino with subsequent leptonic decay 
is performed as a ``blind'' analysis in the final
state with three leptons and significant missing transverse energy. The analysis is carried out in 
224-346 pb$^{-1}$ of data collected by the CDF experiment during the Tevatron Run II.
In the selected events, the prediction of the Standard Model background is in agreement with the observed data.
No evidence of chargino and neutralino production at the kinematically allowed masses is claimed.


\end{document}